\documentclass[12pt]{article}
\usepackage{amsmath}
\usepackage{graphicx,psfrag,epsf}
\usepackage{enumerate}
\usepackage{natbib}
\usepackage{url} 
\usepackage{amsthm}
\usepackage{listings}
\usepackage{moreverb,url}
\usepackage[colorlinks,bookmarksopen,bookmarksnumbered,citecolor=blue, urlcolor=black, linkcolor=blue]{hyperref}
\usepackage{amsmath,amssymb,mathtools,xcolor}
\usepackage{hyperref}
\usepackage{setspace}
\usepackage{fancyvrb} 
\usepackage{graphicx} 
\usepackage{thumbpdf,lmodern}
\usepackage{framed}
\usepackage{upgreek} 
\usepackage{multirow}
\usepackage{booktabs}
\usepackage{kantlipsum}

\newcommand{\blind}{1}

\newtheorem{prop}{Proposition}

\begin{document}

\def\spacingset#1{\renewcommand{\baselinestretch}%
{#1}\small\normalsize} \spacingset{1}


\if1\blind
{
  \title{\bf A Note on Bayesian Modeling Specification of Censored Data in {JAGS}}
  \author{
  	Xinyue Qi \\
  	The University of Texas MD Anderson Cancer Center \\
    and \\
	Shouhao Zhou\thanks{The corresponding author: szhou1@phs.psu.edu}\hspace{.2cm} \\
	Pennsylvania State University \\
	and \\
    Martyn Plummer \\ 
    University of Warwick}
\date{}
\maketitle
}\fi

\if0\blind
{
  \bigskip
  \bigskip
  \bigskip
  \begin{center}
    {\LARGE\bf A Bayesian Analysis of Censored Data in {JAGS}}
\end{center}
  \medskip
} \fi

\bigskip
\begin{abstract}
Just Another Gibbs Sampling ({JAGS}) is a convenient tool to draw posterior samples using Markov Chain Monte Carlo for Bayesian modeling. However, the built-in function \textsf{dinterval()} to model censored data misspecifies the computation of deviance function, which may limit its usage to perform likelihood based model comparison. 
To establish an automatic approach to specify the correct deviance function in {JAGS}, we propose a simple alternative modeling strategy to implement Bayesian model selection for analysis of censored outcomes. The proposed approach is applicable to a broad spectrum of data types, which include survival data and many other right-, left- and interval-censored Bayesian model structures.
\end{abstract}

\noindent%
\vfill

\newpage
\spacingset{1.45} 

\section[Introduction]{Introduction} 
\label{sec:intro}
Censored data are commonly observed in different disciplines such as economics, engineering and life sciences  \citep{lewbel2002nonparametric, hamada1991analysis, chen2012interval}. 
Given the uncertainty in censored data, the modeling and analysis fit naturally in the Bayesian framework by using expectation–maximization (EM), data-augmentation (DA) and Markov chain Monte Carlo  (MCMC) algorithms \citep{dempster1977maximum, tanner1987calculation}.
For example, in highly fractionated experiments, frequentist likelihood-based estimates may not even exist for simple models consisting of only main effects, while Bayesian approach offers a straightforward implementation strategy  \citep{hamada1995analysis}.   
When the outcome cannot be fully observed, censored data can be treated as additional parameters from a fully Bayesian perspective, with a likelihood function specifying joint modeling for both observed and censored data. The Bayesian approach has multiple advantages in the presence of censored data or inadequate sample size, and for nested/non-nested model comparisons \citep{ibrahim2013bayesian}. Compared with multiple imputation, Bayesian modeling is robust in statistical inference even when a large proportion of missing data is present \citep{jakobsen2017and, XQi2020UTSPH}.

Just Another Gibbs Sampling  ({JAGS}) is an object-oriented software to generate posterior samples using MCMC simulations \citep{plummer2003JAGS}. It simplifies the implementation of Bayesian modeling by only requiring the specification of likelihood functions and prior distributions, making it unnecessary to specify the conditional distributions for model parameters, especially when the closed form expressions are not available. {JAGS} also clarifies certain confusing aspects for missing data in {BUGS} \citep{spiegelhalter2003winbugs, lunn2012bugs}. To distinguish the concepts of censoring and truncation, it introduces a degenerate \textsf{dinterval}  distribution function for general interval-censored data \citep{plummer2003JAGS}.

Some existing \texttt{R} packages, including \texttt{rjags} \citep{pkg:rjags}, \texttt{r2jags} \citep{su2015package} and \texttt{runjags} \citep{denwood2016runjags}, provide a user-friendly interface for R users to conduct Bayesian data analysis via {JAGS}. Most importantly, these  R packages for {JAGS}, together with \texttt{coda} \citep{plummer2006coda} and \texttt{MCMCpack} \citep{martin2011mcmcpack}, not only make it easy to process the output of Bayesian models implemented using JAGS, but also further help (1) visualize the posterior samples via plots, (2) predict new data based on posterior predictive distributions, and (3) calculate the deviance using posterior samples from {JAGS} models.

For Bayesian inference especially with complicated model features, model selection is a critical component to identify an approximate model best describing the information in the data. Among many popular approaches, the seminal work of deviance information criterion (DIC) by \citet{spiegelhalter2002bayesian} was proposed based on Kullback-Leibler (K-L) divergence \citep{kullback1951information} 
and embedded in {JAGS} as part of the \textsf{dic} module based on the posterior samples obtained from MCMC simulations. However, when the outcome variables are censored, the built-in function \textsf{dinterval()} returns a constant value of 1 for the likelihood calculation \citep{kruschke2014doing, plummer2017JAGS}, which is equivalent to ignoring all of the censored observations in the deviance monitor of the \textsf{dic} module. As a result, it fails to calculate DIC for model comparison, which may limit the broader usage of {JAGS} for Bayesian modeling of censored data \citep{sourceforge2012}.

Therefore, we propose an alternative model specification for analysis of censored outcomes in {JAGS}.  It is a universal approach that automatically returns the correct deviances for both observed and censored data, such that DIC and penalized expected deviance \citep{plummer2008penalized} can be properly and simultaneously calculated using posterior samples from MCMC simulations; thus Bayesian model selection for censored data modeling can be conducted using {JAGS} without analytical customization of the deviance of the model. The proposed approach is applicable to many different Bayesian model structures, such as Bayesian tobit regression model \citep{chib1992bayes}, semiparametric accelerated failure time (AFT) models for censored survival data \citep{ghosh2006semiparametric}, illness-death model using Bayesian approach for semicompeting risks data \citep{han2014bayesian}, Bayesian hierarchical model for censored normal outcome \citep{carvajal2017virus}, and Bayesian Thurstonian models for ranking data \citep{johnson2013bayesian}, among many. 

The rest of the paper is organized as follows. The default approach for censored data modeling using built-in function in {JAGS} is introduced in Section  \ref{sec:current}. The alternative strategy for correct deviance computation is proposed in Section  \ref{sec:alter}. In Section \ref{sec:examples}, we use a right-censored survival example to illustrate the discrepancy in deviance functions using both approaches, and applied Bayesian model selection using the correctly specified likelihood in an application to drug safety for cancer immunotherapy. Concluding remarks and discussions are given in Section \ref{sec:dis}. 

\section[Default procedure for censored data modeling in JAGS]{Default procedure for censored data modeling in {JAGS}} \label{sec:current}

Censoring occurs when the value of an observation is only partially observed, which is common in Bayesian modeling. We first briefly review the standard approach to model censored data in {JAGS} with its limitation in model assessment.

A default approach for analysis of censored observations in {JAGS} is to use the built-in \texttt{dinterval} distribution function for model specification and posterior sampling.  The \textit{Model 1} below illustrates a general form of model specification for censored data analysis in {JAGS}. It helps modeling three types of censoring: right-censoring, left-censoring and interval-censoring \citep{plummer2017JAGS}. 
\begin{Verbatim}[tabsize=6] 
model{ # Model 1
	for (o in 1:O){ # O is the number of observed cases;
		Y[o] ~ f(theta[o])  # f need to be specified for JAGS
	} 

	for (j in 1:J){ # J is the number of censored observations;
		# Left censoring (R=0): lim[j,] = c(cut[j], inf); 
		# Right censoring (R=2): lim[j,] = c(-inf, cut[j]); 
		# Interval censoring (R=1): lim[j,] = c(cut1[j], cut2[j]);
		R[j] ~ dinterval(Y[j], lim[j,]) 
		Y[j] ~ f(theta[j])
	}
	
	# prior for theta's 
}
\end{Verbatim}

where the outcome of interest, $Y$, which can be either observed or censored (coded as $\texttt{NA}$ in the data table), follows density distribution \texttt{f} with parameter $\theta$. $R$ is a censoring variable following an interval distribution. If $R=1$, then the outcome is interval-censored; If $R=0$, the data is left-censored while outcome contains partial information which is less than a lower limit; If $R=2$, the data is right-censored, which is above a certain cutoff value. \texttt{lim[,]} is a vector of length 2, which contains a pair of cutoff values for each unobserved outcome data, as illustrated in the comment lines above. 

However, \texttt{dinterval()} function has a limitation in deviance calculation when we assess model fit based upon deviance-based statistics. For example, when we apply an existing function, \texttt{dic.samples()}, in the \texttt{rjags} package \citep{pkg:rjags} to call the \texttt{dic} module and to generate penalized deviance samples within {R} \citep{Rprogram}, the following warning message appears. 
\begin{Verbatim}[tabsize=6] 
    Warning message:
	    In dic.samples(model=model, n.iter=n.iter, type="pD"):
		    Failed to set mean monitor for pD
	    Support of observed nodes is not fixed
\end{Verbatim}

By default, the \texttt{dic} module was created to monitor and record the likelihood/deviance of a {JAGS} model at each iteration and calculate the deviance-based model selection criteria such as DIC or penalized expected deviance. In the presence of censored outcomes, even though the \texttt{dinterval()} function can generate the proper posterior distribution of the parameters in {JAGS},  the likelihood function is misspecified with \emph{the wrong focus} of inference on the censored outcome variable \citep{sourceforge2012}. Instead, a constant value of 1 for the likelihood function, or equivalently, a constant value of 0 for the deviance function, is misspecified for the censored outcomes in the deviance monitor. Therefore, the posterior mean deviance computed from the \texttt{dic} module using the default procedure \texttt{dinterval()} is mistakenly reported by the posterior mean deviance of observed data only. It suggests that the posterior mean deviance extracted from the \texttt{dic} module in {JAGS} should not be used in model assessment \citep{kruschke2014doing}. 

\section[Alternative modeling strategy in JAGS]{Alternative modeling strategy in {JAGS}}
\label{sec:alter}

The goal is simply to derive the deviance and associated model selection criteria in {JAGS} without any manual calculation by definition.  Rather than handling censored data with the \texttt{dinterval()} function in the {JAGS} \textit{Model 1}, we present an alternative modeling strategy to specify the proper deviance based on the type of censoring. 

We divide the data into 3 subgroups: observed, left- or right-censored, and interval-censored.  For incomplete observations, we introduce ancillary indicator variables $Z_1$ for left- and right- censored data and $Z_2$ for interval-censored data. Hence, the alternative {JAGS} model specification (\textit{Model 2}) can be written in a general form as follows: 

\begin{Verbatim}[tabsize=6] 
model{ # Model 2
	# block 1: fully-observed
	for (o in 1:O){
		Y[o] ~ f(theta[o])   # f need to be specified  for JAGS
	} 
	
	# block 2: left/right censoring
	for (c in 1:C){
		Z1[c] ~ dbern(p[c]) 
		p[c] <- F(cut[c], theta) 
	}
	
	# block 3: interval censoring
	for (i in 1:I){
		Z2[i] ~ dbern(p[i])
		p[i] <- F(cut2[i], theta) - F(cut1[i], theta) 
	}
	
	# prior for theta's
}
\end{Verbatim} 

Every subgroup is self-blocked with a separate section of the likelihood in {JAGS}, where $O$ is the set of observed data, $C$ is the set of left/right-censored observations, and $I$ is the set of interval-censored observations. $Z_1$ is a binary random variable, where $Z_1=1$ if it is left-censored, or $Z_1=0$ if right-censored. The probability of success $p$ in Bernoulli distribution of $Z_1$ is defined by the cumulative distribution $F$ for the censored outcomes, which neatly identifies the probabilities for both left-censored and right-censored data with properly specified cutoffs. For interval censored observations, we set $Z_2=1$ and the probability of success in Bernoulli distribution is the incremental change of the values in $F$ function between the cutoffs, corresponding to the unobserved outcome which lies in a semi-closed interval. 

Proposition \ref{prop} in \hyperlink{sec:technical1}{\emph{Appendix A}} demonstrates that the proposed alternative modeling strategy in the {JAGS} \textit{Model 2} has a correctly specified likelihood function for censored data. Therefore, it is warranted that the {JAGS} \textit{Model 2} can generate proper posterior samples and deliver valid Bayesian posterior inference. 

In addition, the {JAGS} \textit{Model 2} automatically specifies correct deviances in the \texttt{dic} module for model assessment of censored observations. For K-L based model comparison, especially when there are complicated model features, it is convenient to have an automatic algorithm to avoid any manual calculation of deviance function and model selection criteria. Because the computation is implemented \textit{via} the built-in \texttt{dic} module, we empirically compare the deviance reported from the {JAGS} \textit{Model 2} to the deviance manually calculated using posterior samples in the next Section and illustrate that the proposed model can report the correct deviance values.

The {JAGS} \textit{Model 2} encompasses a broad range of model structures. The censored regression models, which are also called tobit models, usually have data both in blocks 1 and 2 with normally distributed or \textit{t}-distributed errors \citep{chib1992bayes, long1997regression}. Some extensions include time-series analysis \citep{lee1999estimation}, longitudinal data analysis \citep{twisk2013applied} and spatial analysis \citep{xu2015maximum}. In the context of survival data analysis, some commonly assumed parametric distributions $F$ include exponential, Weibull, generalized gamma, log-normal, and log-logistic \citep{klein2006survival, kalbfleisch2011statistical}, since the event times are positively valued with a skewed distribution, making the symmetric normal distribution a poor choice for fitting the data closely. Additionally, it is unnecessary to assume a known censoring time. Because the cutoff can be either pre-specified with a fixed value or modeled as a random variable, the proposed approach naturally accommodates models with unobserved, stochastic censoring thresholds \citep{nelson1977censored}.  

Even for non-censored data, the proposed modeling strategy can still be useful in some situations for computational advantages. After converting the standard model to a latent-variable formulation, we can adapt {logit}, {probit} or complementary {log-log} models as a type of block 2 data with $Z_1$ defined as the binary outcome and \texttt{cut} (cutoff) treated as fixed at 0 \citep{freedman2009statistical}. It is also possible to extend the proposed approach for ordered probit analysis \citep{albert1993bayesian}, which accommodates many applications in economics and marketing \citep{koop2003bayesian}.

\section{Illustrative Examples}
\label{sec:examples}
In this section, two real data applications are examined with the proposed approach. The first example in Section \ref{sec:ex1} applies both the default approach and the alternative strategy to model time-to-event outcome with right censoring. The reported deviance of the model is assessed with the true value calculated manually based on the full likelihood function.  It demonstrates that the alternative strategy not only properly draws posterior samples in {JAGS}, but also automatically delivers the correct deviance for model assessment.  The second example in Section \ref{sec:ex2} shows that the proposed approach is capable of comparing censored data models by DIC \citep{spiegelhalter2002bayesian} and penalized expected deviance (PED, \citealt{plummer2008penalized}) simultaneously, using a drug safety subset \citep{wang2019treatment} in which some of the outcome data are missing not at random (MNAR).

\subsection{survival data} 
\label{sec:ex1}
Right censoring is common in the time-to-event data of survival analysis. The first example is from a classical right-censored survival dataset on acute myeloid leukemia \citep{miller2011survival}. Individual patient-level data were collected along with survival or censoring time to test whether the standard course of chemotherapy should be maintained for additional cycles or not. The Bayesian survival analysis is conducted using MCMC simulation and implemented in {JAGS} 4.3.0 software \citep{plummer2017JAGS} and \texttt{R} version 3.4.1. The {JAGS} codes for both models are attached in \hyperlink{sec:technical2}{\emph{Appendix B}}. We run three parallel chains for the model and discard 30,000 iterations of burn-in, followed by 10,000 posterior samples of hazard rates per MCMC chain with thinning in the exponential survival regression model. Once the posterior samples are obtained, the deviance function of the model based on the exact likelihood function is manually calculated, and compared with the calculated deviance using \texttt{dic.samples()} function in the \texttt{rjags} package with additional 10,000 iterations. 
\begin{figure}[h!]
	\centering
	\includegraphics[width=1.0\linewidth,angle=90,scale=0.38]{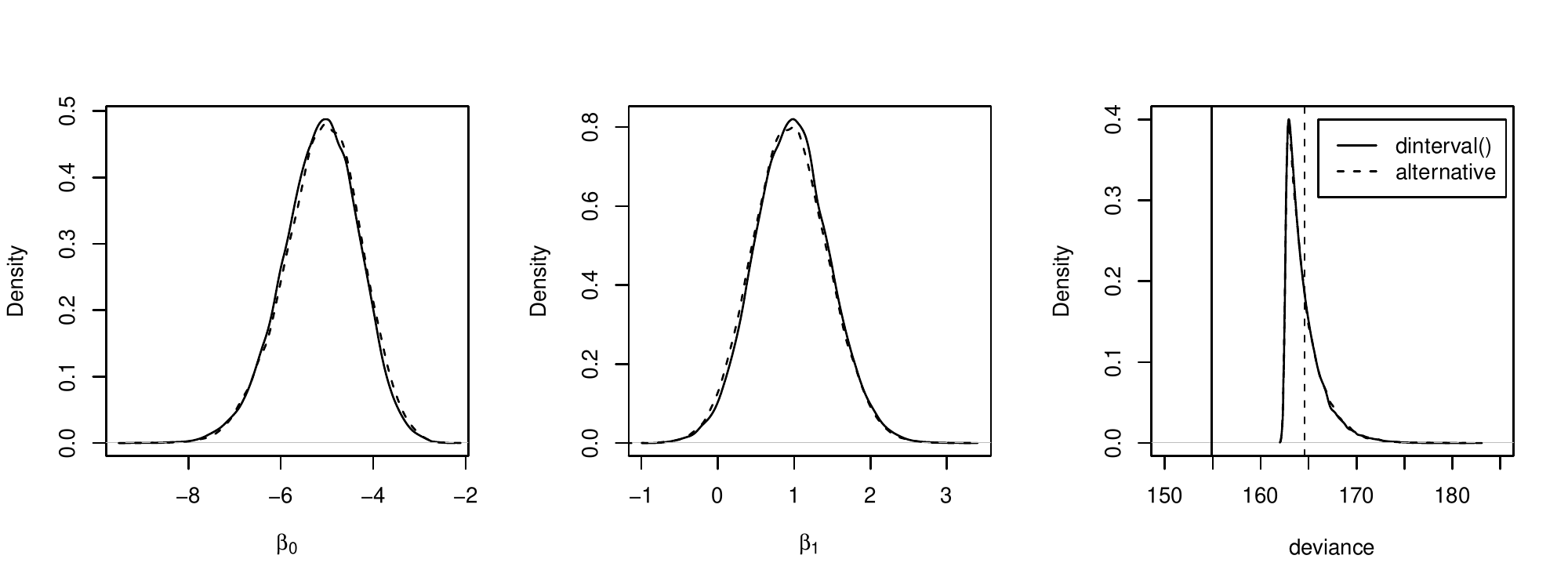}
	\caption{(a): A kernel density plot of regression coefficient $\beta_0$ (the log of the baseline hazard) in the exponential survival regression model comparing two methods; (b): A kernel density plot of regression coefficient $\beta_1$ (the log of the hazard ratio in patients who maintain additional cycles of chemo relative to patients who do not) comparing two methods; (c):  A kernel density plot of deviance functions comparing two methods by manual computation of deviance from posterior samples (based upon the exact likelihood). The two vertical lines show the mean deviances generated via the \texttt{dic.samples()} function by the two methods.}
	\label{fig:1}
\end{figure} 

Figure \ref{fig:1}(a) on the left and \ref{fig:1}(b) in the middle compare the kernel density plots of posterior samples for coefficients in the exponential survival regression model between the default approach using \texttt{dinterval()} and the alternative strategy. The proposed approach has almost identical distribution to the default approach using \texttt{dinterval()} in estimation of the coefficient parameters. The output of \texttt{dic.samples()} function for mean deviance estimation is plotted in Figure \ref{fig:1}(c) on the right, where the solid vertical line shows the mean deviance using \texttt{dinterval()} function and the dashed vertical line using the proposed alternative strategy. Based on 30,000 posterior samples of each method, we also manually calculate the deviance based on the exact likelihood (\ref{likelihood function 1}) and plot their kernel density curves displayed in the last panel. The result demonstrates that the proposed {JAGS} \textit{Model 2} provides the correct value of mean deviance, while the estimate using \texttt{dinterval()} function is significantly biased due to  the deviances ignored for censored outcomes.

\subsection{binomial data}
\label{sec:ex2}
The second example is from an application to assess drug safety for cancer immunotherapy, known as programmed cell death protein 1 (PD-1) and programmed death-ligand 1 (PD-L1) inhibitors. 
In clinical practice, it is important to investigate the incidences of treatment-related adverse events (AEs) and to better understand the safety profiles of these immuno-oncology drugs. 
In this illustrative example, we apply the alternative strategy after extracting all-grade pneumonitis (a specific type of AE for inflammation of lung tissue) data from a recent meta-analysis \citep{wang2019treatment}. 
The primary response is a binomial outcome for the number of pneumonitis cases that could be censored; some rare pneumonitis data may be missing due to low incidence. Usually, the less frequently observed AEs are less likely to be disclosed, given the prevalent manuscript word count limitations for clinical trial publications in medical journals. For each censored AE, a study-specific cutoff value can be identified; only the AEs either of special interest or with observed incidence exceeding the cutoff were reported. To take those non-ignorable censored data into account, we considered study-level rare binomial AE outcome data within the data coarsening framework \citep{heitjan1991ignorability} to examine the impact of stochastic censoring mechanism. If the data are coarsened at random, then we can construct the resultant likelihood ignoring the coarsening mechanism and model the outcome data only, as is presented below. The complete likelihood can be represented and modeled using \textit{selection model} factorization including sensitivity analysis \citep{little2019statistical}. More technical details can be found in \citet{XQi2020UTSPH}.  

In the Bayesian context, we compare seven distinct censored binomial models for all-grade pneumonitis data to examine the model performance using the proposed strategy. To apply the {JAGS} \textit{Model 2}, an outcome variable $Z_1$ is incorporated for censoring status in block 2. In Model A, a baseline beta-binomial model by complete pooling is to estimate the overall incidence of AE, in which no additional effect is included. In Model B, two-group drug effect is incorporated into the baseline model, and then we can estimate the AE incidences for two drug groups (PD-1 vs. PD-L1 inhibitors). To allow for five drug-specific (Nivolumab vs. Pembrolizumab vs. Atezolizumab vs. Avelumab vs. Durvalumab) effect on the incidence of AE, we begin with modeling drug effects without any link function as Model C, and then extend to specify half-Cauchy prior \citep{gelman2006prior}
to the standard deviation of drug effect with {logit}, {cloglog}, and {probit} link functions in Model D-F, respectively. Lastly, we include a saturated model G to estimate the incidence rate corresponding to each study without pooling. Mean deviance ($\bar{D}$), effective number of parameters ($p_D$), DIC, optimism ($p_{opt}$), and PED are all calculated and compared based on the seven candidate models described above. The model assessment results obtained from the proposed {JAGS} models are summarized in Table  \ref{Table1}. 

\begin{table}[!ht]
	\centering
	{
		\begin{tabular}{@{}cccccccc@{}}
			& Model & $\bar{D}$ & $p_D$ & DIC & $p_{opt}$ & PED & \\ 
			\hline
			& A & 380.85 & 0.99 & 381.84 & 2.05 & 382.90 & \\    
			& B & 371.11 & 1.99 & 373.10 & 4.26 & 375.37 & \\    
			& C & 343.14 & 4.61 & 347.75 & 10.65 & 353.79 & \\  
			& D & 343.35 & 4.56 & 347.91 & 11.02 & 354.37 &  \\  
			& E & 343.39 & 4.54 & 347.93 & 13.19 & 356.58 & \\  
			& F & 343.38 & 4.61 & 347.99 & 10.28 & 353.66 & \\  
			& G & 269.30 & 94.60 & 363.90 & 865.69 & 1134.99 &  
	\end{tabular}} 
	\caption{Model Comparison: posterior mean deviance ($\bar{D}$), effective number of parameters ($p_D$), deviance information criterion (DIC), optimism ($p_{opt}$), and penalized expected deviance (PED) from modeling observed and censored all-grade AE (pneumonitis) data.
	$DIC = \bar{D} + p_{D} $, $ PED = \bar{D} + p_{opt}$. \label{Table1}}
\end{table} 

Per the results summarized in Table \ref{Table1}, there is no significant discrepancy on either DICs or PEDs among Model C-F, indicating that the data are not sensitive to the choice of link functions. In general, models with drug-specific effects (Model C-F) outperform the baseline model with complete pooling (Model A) and the model with PD-1/PD-L1 effect (Model B); the beta-binomial model without pooling (Model G) overfits the data. All results are simultaneously computed from \texttt{dic.samples()} function in the \texttt{rjags} package from {R}.

\section{Discussion} 
\label{sec:dis}

In this paper we propose an alternative strategy to apply Bayesian modeling for censored data in {JAGS}. It specifies the correct deviances for censored observations such that the model selection methods DIC and PED can be easily calculated from the built-in \texttt{dic} module. The proposed approach can also simplify the calculation of other popular Bayesian K-L based measures such as the Bayesian predictive information criterion (BPIC, \citealt{ando2007bayesian}) and the widely applicable information criterion (WAIC, \citealt{watanabe2010asymptotic}). Though not explicitly specified, the proposed approach can be easily extended to model truncated data, for example, left-truncated right-censored observations in survival analysis. Even for non-censored data such as binary outcomes, the proposed approach can still be useful for computational advantages. 

The proposed method may have a similar model presentation to the EM algorithm \citep{dempster1977maximum} to handle censored data, for example, in tobit or probit regression modeling \citep{bock1981marginal, liu1998parameter}. In Bayesian contexts, the EM-type algorithms are designed to apply parameter optimization in the posterior mode estimation, while the goal is to achieve the automatic calculation of deviance with the posterior distribution estimation. DA is another relevant approach to estimate the posterior distribution, which constructs computationally convenient iterative sampling via the introduction of unobserved data or latent variables \citep{tanner1987calculation, chib1992bayes, albert1993bayesian}. Different from our approach, it requires the sampling of the unobserved data, which may alter the deviance in application of K-L based model selection \citep{spiegelhalter2002bayesian}. 

Censoring is frequently observed in real-world data analysis. In addition to normally distributed data in censored regression models, various types of outcome, including survival data \citep{ibrahim2013bayesian}, binomial data \citep{wang2019treatment}, count data \citep{de2017random} and ranking data \citep{johnson2013bayesian}, can all be modeled by the proposed alternative strategy when censoring occurs.  Not only to the medical sciences, the proposed strategy can also be applied to many other fields, such as, in measuring the performance of timing asynchronies using censored normal sensorimotor synchronization data in behavioral science \citep{baaaath2016estimating}, comparing industrial starch grain properties with ordered categorized data in agriculture \citep{onofri2019analysing},  exploring forest genetics by modeling censored growth strain data for narrow-sense heritability estimation in environmental science \citep{davies2017heritability}, determining the importance of influential factors to lower the risk of food contamination for censored microbiological contamination data in food science \citep{busschaert2011hierarchical}, and modeling the interval-censored as well as right-censored time to dental health event in primary school children for public health science \citep{wang2013bayesian}. In summary, the proposed {JAGS} \textit{Model 2} can encompass a broad range of popular model structures and be utilized in a wide spectrum of applications. 

\newpage
\section*{\hypertarget{sec:technical1}{Appendix A}: Alternative Modeling Strategy} 

We justify that the proposed alternative procedure constructs the correct likelihood function for censored outcomes. In likelihood-based inference, the full likelihood for observed and censored data comprises four key components: observed case, left-censored case, right-censored case and interval-censored case. For observed data, the likelihood is simply a product of individual probability density/mass function of observed outcome. For any type of censored cases, the likelihood can be presented in a form of $F_{Y}(b)-F_{Y}(a)$, defining the probability of a censored outcome $Y$ observed in the semi-closed interval, $(a,b]$. Here, $F_Y(y)=\text{P}(Y\leq y)$ denotes the cumulative distribution function of the random outcome variable if it is {fully observed}.  If the outcome variable is left-censored at a cutoff, $y_l$, then $F_{Y}(b) = F_{Y}(y_l)$ and $F_{Y}(a) = F_{Y}(-\infty) = 0$. If data is right-censored with a lower bound, $y_r$, then $F_{Y}(a) = F_{Y}(y^-_r)$ and $F_{Y}(b) = F_{Y}(+\infty)= 1$. For interval-censored data, the likelihood function is the product of $\text{Pr}(u_i \le Y \le v_i) = F_Y(v_i) - F_Y(u^-_i)$, where $u_i$ and $v_i$ are a pair of interval thresholds, which could vary for every observation. Therefore, the exact likelihood function is given by:
\begin{equation}
	\mathcal{L}_{exact}\left (\theta; y \right ) = \prod_{o \in O}f_Y\left (y_o \right ) \prod_{l \in L}F_Y\left (y_l \right )
	\prod_{r \in R} \left [1-F_Y\left (y^-_r \right )\right ] 
	\prod_{i \in I} \left [F_Y\left (v_i \right )-F_Y\left (u^-_i \right )\right ],
	\label{likelihood function 1}
\end{equation}
where $O$ is the set of observed outcome, $L$ (or $R$) is the set of left (or right) censored observations, and $I$ is the set of interval-censored data with $u_i$ and $v_i$ denoting the lower and upper bound of the $i${th} interval-censored observation. 

In the {JAGS} \textit{Model 2}, we can specify the cutoff value \texttt{cut} = $y_l$ if data are left-censored, \texttt{cut} = $y^-_r$  if data are right-censored, and (\texttt{cut1}, \texttt{cut2}) = $(u^-_i, v_i)$ if data are interval censored. Defining \texttt{F} = $F_Y$, we have the following property for the likelihood from the proposed {JAGS} model.

\begin{prop} 
	The likelihood generated from the {JAGS} \textit{Model 2} using Bernoulli distribution with the cumulative probabilities for censored data is identical to the exact likelihood (\ref{likelihood function 1}). 
	\label{prop}
\end{prop}

\vspace{-.1in}
\begin{proof}
To illustrate that the likelihood from the {JAGS} \textit{Model 2}, $\mathcal{L}_{jags}$, is identical to its exact likelihood, $\mathcal{L}_{exact}$,  we start with deriving the formula for the likelihood presented in the censored {JAGS} model, which has three major components: observed case, one-sided censored case, and interval-censored case. 
The full likelihood, $\mathcal{L}_{jags}$, can be written as:
\begin{equation}
	\begin{split}
	\mathcal{L}_{jags}\left (\theta; y \right ) &= \prod_{o \in O}f_Y\left (y_o \right ) 
	\prod_{c \in C} \left \{ \left [ F\left (\texttt{cut}_{c} \right ) \right ]^{I\left (Z_{1,c}=1 \right )}
	\left [1-F\left (\texttt{cut}_c \right )\right ]^{I\left (Z_{1,c}=0 \right )} \right \} \\
	&\quad \prod_{i \in I} \left [F\left (\texttt{cut2}_i \right )-F\left (\texttt{cut1}_i \right )\right ]^{I\left (Z_{2,i}=1 \right )} \\
	&= \prod_{o \in O}f_Y\left (y_o \right )
	\prod_{\substack{c \in C \\ \left \{ Z_{1,c}=1 \right \}}} F\left (\texttt{cut}_{c} \right )
	\prod_{\substack{c \in C \\ \left \{ Z_{1,c}=0 \right \}}} \left [1-F\left (\text{cut}_{c} \right )\right ] \\
	& \prod_{\substack{i \in I \\ \left \{ Z_{2,i}=1 \right \}}} \left [F\left (\texttt{cut2}_{i} \right )-F\left (\texttt{cut1}_{i} \right )	\right ] \\
	&= \prod_{o \in O} f_Y\left (y_o \right ) \prod_{l \in L} F_Y\left (y_l \right )
	\prod_{r \in R} \left [1-F_Y\left (y^-_r \right )\right ]
	\prod_{i \in I} \left [F_Y\left (v_i \right )-F_Y\left (u^-_i \right )\right ].
	\label{likelihood function 2}
	\end{split}
\end{equation}
\end{proof}

\section*{\hypertarget{sec:technical2}{Appendix B}: JAGS code for survival example}

The following is the {JAGS} code for survival regression model in Section \ref{sec:ex1}.

\begin{Verbatim}[tabsize=6]
# The default approach implemented in JAGS
model{
	for (j in 1:J){
		R[j] ~ dinterval(Y[j],lim[j]) # right-censored
		Y[j] ~ dexp(lambda[j])
		lambda[j] <- exp(b0+b1*group[j])
	}
	b0 ~ dnorm(0, tau0) # tau0 fixed at 0.01
	b1 ~ dnorm(0, tau1) # tau1 fixed at 0.01
}
		
# The proposed approach implemented in JAGS 
model{
	for (o in 1:O){
		Y[o] ~ dexp(lambda.adj[o]) # observed
		lambda[o] <- exp(b0 + b1*group[o])
	}
	for (c in 1:C){
		Z[c] ~ dbern(p.adj[c])              # censoring status				
		p[c] <- pexp(cut[c],lambda[c+O])    # cumulative exp. dist.
		lambda[c+O] <- exp(b0 + b1*group[c+O])
	}
	b0 ~ dnorm(0, 0.01)
	b1 ~ dnorm(0, 0.01)
}	
\end{Verbatim}

\newpage
\bibliographystyle{Chicago}
\bibliography{final}

\end{document}